\documentstyle[psfig,proceedings]{crckapb} 
\newcommand{\philband}{$\rm A^{1}\Pi_{u}-X^{1}\Sigma^{+}_{g}$}
 
\newcommand{\redband}{$\rm A^{2}\Pi-X^{2}\Sigma^{+}$} 
\newcommand{\sysch}{$\rm A^{1}\Pi-X^{1}\Sigma^{+}$}

\begin{opening}
\title{LINES OF CIRCUMSTELLAR C$_{2}$, CN, and CH$^{+}$ IN THE OPTICAL 
SPECTRA OF POST-AGB STARS}

\author{ERIC J. BAKKER~} \author{~DAVID L. LAMBERT} \institute{Department of
Astronomy and McDonald Observatory\\ University of Texas, Austin TX,
U.S.A.} \author{EWINE F. VAN DISHOECK}
\institute{Leiden Observatory, University of Leiden, The Netherlands}
\runningauthor{E.J. Bakker, ~D.L. Lambert, ~and ~E.F. van Dishoeck} 
\runningtitle{OPTICAL SPECTRA OF CIRCUMSTELLAR MOLECULES}
\end{opening}

\begin{document}

\begin{abstract}
Recent optical spectra of post-AGB stars show the presence~of C$_{2}$, CN,
and CH$^{+}$ originating in the circumstellar shell.  We present here new,
higher resolution spectra which provide constraints on the physical
para\-meters and information on the line profiles. An empirical curve of
growth for the C$_{2}$ Phillips and CN Red system lines in the spectrum of
HD~56126 yields $b = 0.50^{+0.59}_{-0.23}$ km s$^{-1}$.
CH$^{+}$ (0,0) emission lines in the spectrum of the Red Rectangle have been
resolved with a FWHM $\approx 8.5 \pm$ 0.8 km\,s$^{-1}$.  The circumstellar CN
lines of IRAS~08005--2356 are resolved into two separate components with
a velocity separation of $\Delta v = 5.7 \pm 2.0$ km s$^{-1}$. The line 
profiles of CN of HD~235858 have not been resolved.

\end{abstract}
\vspace*{-0.5cm}

\section{Introduction}

Post-AGB stars are in a transition stage between the Asymptotic Giant
Branch (AGB) and the planetary nebulae (PN) stage. 
During the early stage of
post-AGB evolution the star is obscured by material expelled during the AGB
phase (the AGB ejecta). As this ejecta slowly moves away from the central
star, the optical depth decreases and the star can be detected in the
optical. When the star reaches high enough temperatures, the AGB ejecta is
ionized and is observable as a planetary nebula.

We have studied optically bright post-AGB stars (spectral type A to G
supergiants) which show circumstellar molecular line absorption (C$_{2}$
and CN, or CH$^{+}$) or emission (CH$^{+}$) in their optical spectra.  The
radial velocities and low excitation temperatures of the molecules
(Bakker et al. 1996a,b) 
identify them as circumstellar rather than photospheric or interstellar
(see also Hrivnak 1995). The excitation of C$_{2}$ is 
generally described by rather high temperatures, $T_{\rm ex}=43-399$ K,
whereas the CN excitation is found to be much lower, $18-50$ K. The
reason for this difference is discussed in section 2.2. The observed
abundances and excitation of the molecules can lead to a 
determination of the mass-loss rate.

We have used the C$_{2}$ (\philband) and CN (\redband) data of HD~56126
to study optical depth effects by means of  the curve of growth,
and we present the first results of a survey to observe these optical
molecular bands at high spectral
resolution ($R\geq 120\,000$). Our primary
goal is to resolve the line profiles and to determine the Doppler
parameter $b$ 
and the chemical (e.g. abundances)
and physical (e.g. expansion velocities and temperatures)
conditions of the circumstellar shell.

\section{Curve of Growth Analysis for HD~56126}

\subsection{The empirical curve of growth}

\begin{figure*}
\centerline{\hbox{\psfig{figure=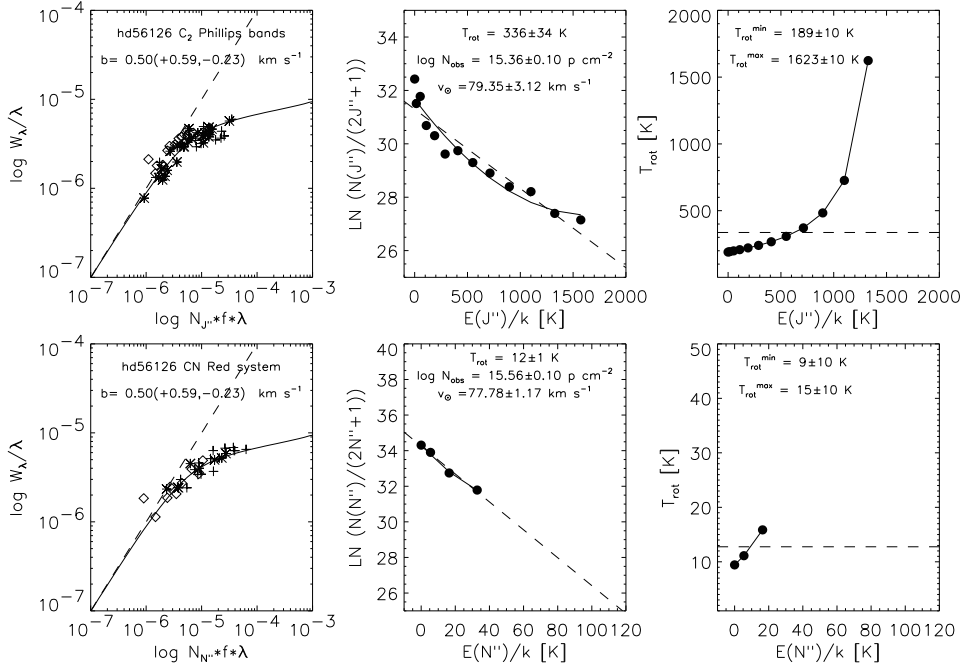,width=\columnwidth}}}
\caption{Curve of growth analysis for three  C$_{2}$ Phillips and
three  CN Red system bands in the optical spectrum of HD~56126
(diamond: $v'=3$, asterisk: $v'=2$, plus: $v'=1$, all for $v''=0$).
The upper three panels concern C$_{2}$ and the lower three panels
concern CN.
Left panels: CoG with the best theoretical CoG over-plotted.
Middle panels: optical-depth-corrected rotational diagram
with a fit to the data assuming a Boltzmann distribution 
(dashed line) and a smooth non-Boltzmann distribution
(solid line).
Right panels: 
the rotational temperature determined from two successive energy levels
(the dashed line is the average $T_{\rm rot}$).}
\end{figure*}

A curve of growth (CoG) has been empirically determined for the 
$^{12}$C$^{12}$C Phillips (\philband)
($v'=1,2,3$, $v''=0$, $J''\leq 24$) 
and the  
$^{12}$C$^{14}$N Red system (\redband)
bands ($v'=1,2,3$, $v''=0$, $N''\leq 3$) of HD~56126 (Figure 1).
Our motive for this investigation is to  decide whether the assumption
of optically thin lines for the weaker bands (e.g. 3,0) is allowed.
The equivalent widths are taken from the work
of Bakker et al. (1996a,b). The range of line oscillator strengths is 
$6.67 \times 10^{-5}$ to 
$1.44 \times 10^{-3}$ and 
$7.19 \times 10^{-5}$ to 
$9.79 \times 10^{-4}$ for C$_{2}$ and CN, respectively.

Because three C$_{2}$ Phillips bands have been observed, the 
CoG has for each lower energy level $J''$ up to
nine transitions (a P, Q, and R branch for each band). This
redundancy allows the determination of the CoG.
All transitions from a given $J''$ level  are fitted to the CoG
by changing the column density of that level $N(J'')$.
For the  CN Red system the redundancy
is higher due to spin-doublet splitting of the lower and
upper electronic state 
and $\Lambda$-type doubling of the upper electronic state.
There are twelve transitions from each $N''$ level:
six main and  six satellite branches.
Three bands were used, which gives in total 
36 allowed transitions per $N''$ level ($F$$_{1}$ and $F_{2}$).

After all observed $J''$ (or $N''$) levels have been fitted to the CoG 
(Fig.~1, left panels) an optical-depth-corrected 
absolute rotational diagram (Fig.~1, middle panels)
gives the absolute population for  each
$J''$ (or $N''$) level. Under the assumption
of a Boltzmann distribution, a linear fit to the diagram gives the
average rotational temperature $T_{\rm rot}$. Finally the rotational
temperature from two successive energy levels can be determined as
a function of the lower energy level (Fig.~1, right panels). 
The parameters derived from
this analysis are given in Table~1.

\subsection{Results and interpretation}

The empirical CoG (Fig.~1)
shows clearly that the observations cover the optically-thin and
saturated parts of the CoG.
The theoretical CoG has a Doppler parameter
$b =0.50^{+0.59}_{-0.23}$~km s$^{-1}$ 
and $\tau \approx 1$ is reached at an
equivalent width of $EW ( \tau \approx 1)=18$ m\AA. 
Since the strongest lines in the (3,0) transitions of C$_{2}$ and CN have
equivalent width of 38 and 34 m\AA, respectively, the optically thin
approximation is only valid  for the weaker lines in these bands.
We note that all the lines of the C$_{2}$
($v',v''$) = (3,0) band seem to be offset with respect 
to the other lines. This might indicate that the band oscillator
strength used is somewhat too low.

\begin{table}
\caption{Results from the curve of growth analysis of C$_{2}$ and CN in
the optical spectrum of HD~56126}
\begin{center}
\begin{tabular}{lrrl}
\hline  
                      &C$_{2}$ \philband & CN \redband    &             \\
\hline
$b$                   &$ 0.50(+0.59,-0.23)$&$ 0.50(+0.59,-0.23)$& km s$^{-1}$\\
$T_{\rm rot}$ average &$  336\pm34  $&$   12\pm1         $& K           \\
$T_{\rm rot}$ minimum &$  189\pm10  $&$    9\pm10        $& K           \\
$T_{\rm rot}$ maximum &$ 1623\pm10  $&$   15\pm10        $& K           \\
$\log N_{\rm obs}$    &$15.38\pm0.10$&$15.56\pm0.10      $& p cm$^{-2}$ \\
$v_{\rm helio}$       &$79.35\pm3.12$&$77.78\pm1.17      $& km s$^{-1}$ \\
\hline
\end{tabular}
\end{center}
\end{table}

The right panel of Fig.~1 clearly shows
that the  rotational temperature for C$_{2}$ is not constant. The
molecule is therefore not in local thermodynamic equilibrium 
(LTE) and the population distribution over
the rotational energy levels is non-Boltzmann. Van Dishoeck \& Black (1982)
have shown that interstellar
 C$_{2}$ is radiatively pumped. C$_{2}$ is a homonuclear
molecule and does not have allowed pure rotational or vibrational
transitions and can therefore not cool radiatively: $T_{\rm rot} \ge 
T_{\rm kin}$.
For low $J''$ levels the rotational temperature reaches
the kinetic temperature, while for very high $J''$ levels the rotational
 temperature is expected
to reach the color temperature of the local radiation field.
CN on the other hand can effectively cool:
$T_{\rm rot} \le T_{\rm kin}$. 
Based on the population ratio between the C$_{2}$ Phillips band
$J''=0$ and $J''=2$ levels we 
find: $T_{\rm kin}=189\pm10$ K. Combining this
with the derived $b$ yields $v_{\rm microturb} = 0.34\pm0.6$ km s$^{-1}$.
The measured 
$b = 0.50^{+0.59}_{-0.23}$ km s$^{-1}$ gives  a line profile
with FWHM $= 2 \sqrt{\ln 2} \times b = 0.83$ km s$^{-1}$. In order
to resolve these lines a spectral resolution of $R \geq 360\,000$ is 
needed. In the presence of macroturbulence the lines are broader
and can be resolved at a lower spectral resolution.
The microturbulence is very likely due to a velocity
gradient in the line of sight.

\section{Line Profiles}

\subsection{CH$^{+}$ emission lines of the Red Rectangle}

\begin{figure*}
\centerline{\hbox{\psfig{figure=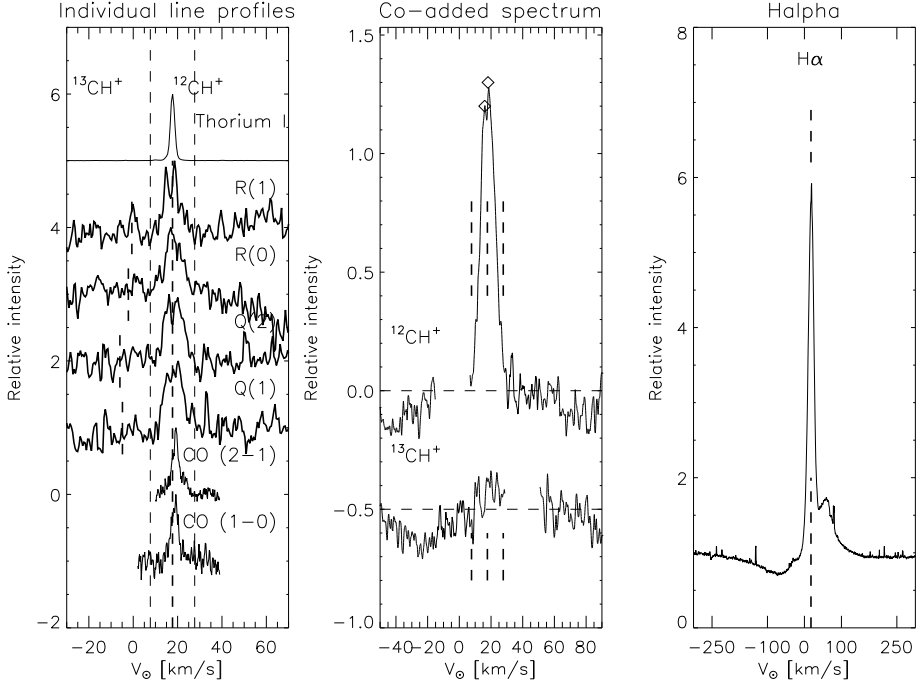,width=\columnwidth}}}
\caption{Left: normalized line profiles of the strongest CH$^{+}$ \sysch~
(0,0)
emission lines and CO radio emission lines (after Jura et al. 1995)
of the Red Rectangle. The thick dashed
lines at 17.7 km s$^{-1}$ 
gives the average velocity, with on both sides (at a
velocity offset of 10.0 km s$^{-1}$) a line
where the intensity of the emission line 
has fallen below the detection limit. The dashed line at approximately
$-0.7$ km s$^{-1}$ ($J''$ dependent) marks
the location where $^{13}$CH$^{+}$ is predicted.
Middle: A non-detection of $^{13}$CH$^{+}$ after the signals
of several lines have been co-added on a velocity axis. It seems that
the $^{12}$CH$^{+}$ lines are a disk-like feature. Right: the H$\alpha$
profile with the central spike at the system velocity.} 
\end{figure*}

The CH$^{+}$ \sysch\/
(0,0) emission band in the optical spectrum of the Red Rectangle
(HD~44179) has been observed at a resolution of $R \approx 120\,000$ using
the 2.7-m telescope of McDonald Observatory (Figure 2).
These lines originate from levels which are 34\,000 K above the ground level.
We have resolved the line profile of the strongest CH$^{+}$ emission lines
and find a FWHM $\approx 8.5\pm0.8$ km s$^{-1}$ (Table~2).  
The intensities of the emission lines fall below the detection limit
for FWFM $\approx 20.0\pm1.0$ km s$^{-1}$.
The line profiles of the R(1), Q(1), and Q(2) lines
suggest a central absorption or the presence of two emission components  
at 16.0 and 18.1 km s$^{-1}$.
The~location of $^{13}$CH$^{+}$ is 
at approximately --0.7 km s$^{-1}$ ($J''$ dependent).
To investigate the presence of weak isotopic lines we have
added the signals of all lines into one single profile (middle panel
of Fig.~2). We have a non-detection of
$^{13}$CH$^{+}$ which yields an isotopic ratio of $^{12}$C/$^{13}$C
$\geq 22$. This is consistent with the progenitor being
a carbon star (Smith \& Lambert 1990).

The line profile of CO is interpreted as due to a broad 
and narrow component (Jura et al. 1995, 1996). 
The broad component has a width comparable to that
of the CH$^{+}$ lines which might suggest that they are formed in the same
region (the circumbinary disk). 
The H$\alpha$ and CO profiles 
shows a strong central emission at the system velocity.
Jura et al. argue that the spike is 
formed in an extended region of ionized gas.

\begin{table}
\caption{Results for the Red Rectangle, HD~235858, and IRAS~08005--2356}
\begin{center}
\begin{tabular}{lllllll}
\hline
                 &Red Rectangle~~  &HD~235858          &IRAS~08005--2356   &\\
                 &CH$^{+}$ (0,0)   &CN Red (2,0)~~     &CN Red (2,0)       &\\
\hline
Date             &1995 Dec. 13     &1995 Sep. 17       &1995 Dec. 12       &\\
HJD              &2450064.7428     &2449977.6607       &2450065.8708       &\\
$v_{\ast, \rm helio}$& $27.7\pm1.4$    &$-34.1\pm0.1      $&$64.5\pm1.0$       &km s$^{-1}$ \\
$v_{\rm mol, hellio}$& $17.7\pm0.1$    &$-48.2\pm0.1      $&$19.4\pm0.2$ (A)   &km s$^{-1}$ \\
                 &                 &                   &$25.2\pm1.0$ (B)   &            \\
$v_{\rm FWHM}   $& $ 8.5\pm0.8$    &$2.2 \pm0.1       $&$~~3.1\pm0.9$ (A)  &km s$^{-1}$ \\
                 &                 &                   &$~~3.4\pm2.0$ (B)  &            \\
$v_{\rm FWFM}   $& $20.0\pm1.0$    &                   &                   &km s$^{-1}$ \\
$^{12}$C/$^{13}$C~~& $\geq 22$     &$\geq 11          $&$\geq 11 $         &\\
\hline
\end{tabular}
\end{center}
\end{table}

\subsection{CN absorption in HD~235858 and IRAS~08005--2356}

HD~235858 has been observed at a resolution of $R \approx 120\,000$.
Figure 3 shows a part of the spectrum which
contains three different categories of  molecular features. The broad
absorption lines are due to photospheric CN ($T_{\rm eff}=5500$ K), 
the narrow absorption lines are circumstellar CN ($T_{\rm eff}$ = 20\,K 
and $\log N = 15.30$ p cm$^{-2}$), and
the three strongest features are telluric H$_{2}$O. 
There is one photospheric atomic line (N\,$\sc{i}$) present in
this spectrum. The circumstellar CN lines are not resolved.
Since Za{\v c}s et al. (1995) did not notice photospheric
CN absorption in their spectrum, it seems that this pulsating
star has only photospheric molecular absorption when the 
stellar effective temperature is sufficient low (when the star
is largest).

\begin{figure*}
\centerline{\hbox{\psfig{figure=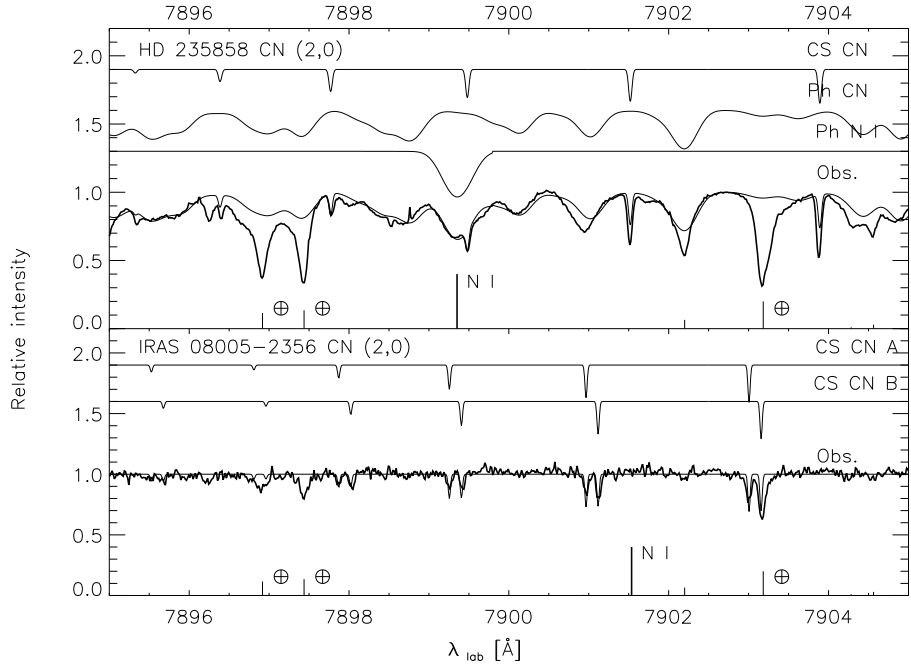,width=\columnwidth}}}
\caption{Upper: a part of the optical spectrum of HD~235858 which
shows the CN Red system (2,0) band lines. Lower: for IRAS~08005--2356,
a synthetic spectrum using photospheric and circumstellar 
CN lines, and N\,$\sc{i}$ (7898.985 \AA) is over-plotted (thin line).
The abscissa is in the rest frame of the molecule.}
\end{figure*}

In an earlier report on the detection of circumstellar CN in
IRAS~08005-2356 we reported the presence of only one broad
component (Bakker et al. 1996b). 
The new spectra (Fig.~3, $R\approx 120\,000$) 
have resolved the broad component into  two separate resolved
components with a velocity separation of $\Delta v = 5.7 \pm 2.0$ km
s$^{-1}$. The two absorption components could be  due
to wind material moving at two different velocities (multiple shells),
or to two photodissociation fronts at different expansion velocities ---
possibly one due to the stellar and one to the interstellar radiation 
field.

\section{Summary}

We have investigated optical depth effect for the circumstellar C$_{2}$
and CN lines in the optical spectrum of HD~56126. From a
curve of growth analysis we find a Doppler parameter 
$b = 0.50^{+0.59}_{-0.23}$ km s$^{-1}$. CN (2,0) lines of
IRAS~08005--2356 and HD~235858 and CH$^{+}$ of the Red Rectangle
have been observed at a resolution $R \geq 120\,000$. CH$^{+}$
has been resolved. The CN lines of IRAS~08005--2356 are resolved into 
two separate resolved components, and the CN of HD~235858 has not been 
resolved.

\end{document}